\documentclass[a4paper]{toledoStyle}
\usepackage{epsfig}

\begin{document}

\title{Submillimeter telescope for the Russian segment of the ISS:
 Submillimetron Project}

\author{V.D.\,Gromov\inst{1,2} \and L.S.\,Kuzmin\inst{2} \and
  D.\,Chouvaev\inst{2} \and L.A.\,Gorshkov\inst{3} \and N.S.\,Kardashev\inst{1} \and
  V.I.\,Slysh\inst{1} \and
  S.F.\,Stoiko\inst{3} \and M.A.\,Tarasov\inst{2,4} \and A.G.\,Trubnikov\inst{1} \and
  A.N.\,Vystavkin\inst{4}  }

\institute{
  Astro Space Center, P.N. Lebedev Physical Institute of the Russian Academy of Sciences (ASC/LPI),
  Moscow, Russia
\and
  Chalmers University of Technology, Gothenburg, Sweden
\and
  Rocket Space Corporation "Energia", Korolev, Russia
\and
  Institute of Radioengineering and Electronics of the Russian Academy of Sciences,
  Moscow, Russia     }

\maketitle

\begin{abstract}

The Submillimetron is the international project of the space telescope
for astronomical studies at the submillimeter and infrared wavelengths
using facilities of the Russian segment of the International Space Station (ISS).
The concept of the telescope includes a 60 cm mirror cooled to liquid helium
temperature with a novel type of microbolometers arrays using effects of superconductivity.
This combination gives unique possibility to realize background limited sensitivity
in the spectral minimum of the extraterrestrial background near frequency 1 THz
between peaks of galactic dust emission and CMB. The angular resolution about 1 arcmin,
field of view about 1\degr, and optics are similar to IRAS satellite, but the sensitivity 
is better on more than order of magnitude for about $10^{-18} W/Hz^{-1/2}$.
This improvement and another spectral region
permits to reveal in full sky survey considerably more new astrophysical objects.
The concept of free flying instrument with periodic docking to ISS gives possibility
to combine low cost with reliability, refilling, repairment and maintenance.

\end{abstract}

\section{introduction}

The initiative of the project (see Gromov, Kardashev, Kurt et al. 2000) was
done in Astro Space Center of the P.N. Lebedev institute RAS after discussions
with NASA and JPL. Detectors are under development in Chalmers University of Technology, Sweden
(Vystavkin et al. 1999, Kuzmin et al. 1999, Gromov et al. 2000). The proposal was
undertaken to feasibility study in S.P. Korolev Rocket Space Corporation Energia and
approved by the Russian Space Agency for the 2-d stage of ISS realization after years 2004 -- 2005.

The size and cooling of the Submillimetron telescope optics is the same as it was in previous
infrared mission IRAS and ISO. Two main features differ it from these missions and new ones
(SIRTF, FIRST, Planck, Astro-F):
\begin{itemize}
  \item free flying spacecraft using facilities of the Russian segment of ISS for deployment
   and maintenance
  \item novel type microbolometers (NHEB) with cooling down to 0.1 K and reaching sensitivity
limited by low extraterrestrial background
\end{itemize}

First feature permits to realize the instrument of relatively low mass and cost,
second one - to surpass new missions submillimeter instruments in sensitivity
for surveys of large sky regions (full sky).

\section{scientific objectives}
\label{gromovv_sec:tit}

First and most bulky goals of Submillimetron observations are:
\begin{itemize}
 \item full sky catalog of submillimeter sources, confusion limited at $1\arcmin$ resolution;
 \item cosmological background study, reducing errors of
 ani\-sotropy measurements by means of more detailed information on foreground sources.
\end{itemize}

In the field of observational techniques the goal is testing of novel type
high-sensitive bolometer, which has not been used for astronomical observation and was
especially designed for conditions of low extraterrestrial background.
The experiment can be also testbed for future large cryogenic telescopes,
including ASC/LPI Millimetron project for Lagrange point L2 (\cite{gromovv:trudyMM}).

Other tasks:
\begin{itemize}
 \item study in submillimeter and infrared wavelengths of the "cold" component of the matter
 in the Universe (dust in the Solar system, Galaxy, and in extragalactic sources)
 \item study of the anisotropy of the cosmic microwave background radiation and search
 for Lyman-alpha line in the epoch of recombination and secondary heating
 \item studies of the spectra of astronomical sources and their variability.
\end{itemize}

Targets of observations:
\begin{itemize}
 \item CMB - cosmological background radiation, spectrum, mapping;
 \item clusters of galaxies, dust, SZ-effect;
 \item "mirror" galaxies, relativistic particles, dust;
 \item IR galaxies, dust, cosmological evolution;
 \item AGN - active galactic nuclei, spectrum, variability;
 \item interstellar dust in our galaxy, spectrum, mapping;
 \item cocoons, young stars, star envelopes, protoplanets;
 \item cold stars, peculiar and variable stars;
 \item neutron stars and galactic black holes, remnants of supernovae;
 \item center of galaxy \object{Sgr~A}*;
 \item  interplanetary dust, belts;
 \item  planets, asteroids and comets;
 \item  Dyson spheres (CETI).
\end{itemize}

\begin{figure}[ht]
  \begin{center}
    \includegraphics[width=7cm]{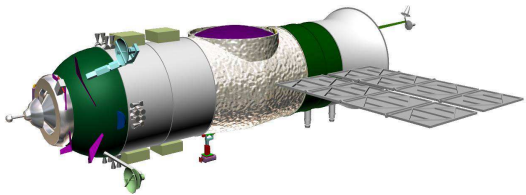}
  \end{center}
\caption{ Free-flying spacecraft for RS/ISS.
At left is docking assembly, antennas of docking system, and pressurized compartment.
Center - nonpressurized compartment for telescope.
At right - aggregates compartment and solar panels. }
\label{gromovv_fig:fig1}
\end{figure}

\section{The mission}
\label{gromovv_sec:cmd}

\begin{figure}[ht]
  \begin{center}
    \includegraphics[width=7cm]{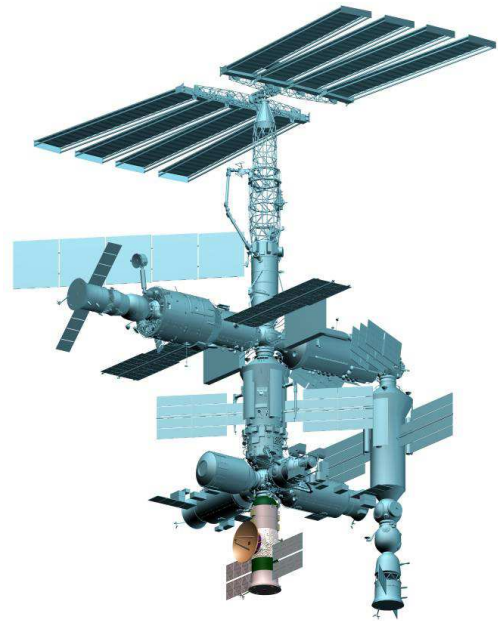}
  \end{center}
\caption{ Submillimetron module docked to Russian Segment of the ISS for service and instruments
replacement. Observational phase begins after screens deployment and detachment from ISS }
\label{gromovv_fig:fig2}
\end{figure}

Design of the free-flying spacecraft (Figure~\ref{gromovv_fig:fig1}) for Submillimetron is 
based on Progress cargo ship and Soyuz rocket. Their reliability was proved in numerous
space flights. Main scientific payload of the spacecraft is cryogenic telescope
for submillimeter photometry in sky survey. A call for proposal on additional
scientific instruments for heterodyne spectrometry and interferometry is now in stage of preparation.
The launch is approved for the 2-d stage of ISS realization, when finished its assembling.
After docking to ISS, deployment of the telescope and screens there will begin a period of free-flying
observation with duration of 1-2 year. In this period docking for repair and replacement of
instruments parts are permitted. After the end of this observation period and docking to
Russian segment of ISS (Figure~\ref{gromovv_fig:fig2}) an upgrade of instrument is planned 
with attachment of larger telescope mirror with diameter 3.5 m or more.

\bigskip

\section{The Instrument}
\label{gromovv_sec:fig}
\begin{figure}[ht]
  \begin{center}
    \includegraphics[width=7cm]{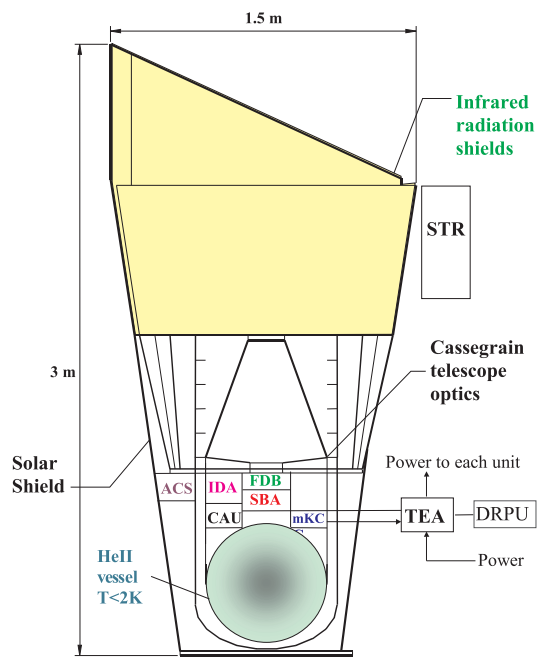}
  \end{center}
\caption{ Schematical view of the cryogenic telescope. \protect\newline
ACS - Active Cooling System, T=20K; \protect\newline
STR - Star Tracker, defines pointing of the telescope; \protect\newline
HeII - superfluid helium vessel T=2K for cooling of \protect\newline
focal instruments, primary and secondary mirrors on orbit \protect\newline
(warm start); \protect\newline
FDB - Focal Dichroic Beam-splitters assembly; \protect\newline
IDA- Infrared Detector Array; \protect\newline
SBA - Submillimeter Bolometer Arrays assembly; \protect\newline
CAU - Cool Amplifiers Unit; \protect\newline
mKC - milli-Kelvin Cooler (T=100 mK); \protect\newline
TEA - Telescope Electronics Assembly; \protect\newline
DRPU - Data Registration and Processing Unit. }
\label{gromovv_fig:fig3}
\end{figure}

Cryogenic telescope of the Submillimetron project consists of Cassegrain optical system entirely cooled
with several stages of refrigeration: radiation cooling, active cooling machine, and passive cooling
with storage of liquid helium. Additional stage of refrigeration included for cooling of bolometers
for subhelium temperatures. Main focal instrument - submillimeter photometers includes bolometers arrays
and dichroic mirrors. A whole design of the telescope and bolometers is optimized for achievement of
sensitivity close to fundamental limit of fluctuations of extraterrestrial background in its spectral
minimum, which takes place in submillimeter region between spectral peak of galactic dust emission and CMB.
Additional focal instruments are infrared photometer and/or submillimeter superheterodyne focal instrument.
Schematical view of the cryogenic telescope is shown in Figure~\ref{gromovv_fig:fig3}, 
main parameters of the instrument - in Table~\ref{gromovv_tab:tab1}.

\begin{table}[hb]
\caption{ Submillimetron telescope parameters }
\label{gromovv_tab:tab1}
\begin{tabular}{ll}
\hline \\
Telescope diameter: & D=0.6 m \\
Field of view & 1\degr \\
Angular resolution: & 1\arcmin-10\arcmin. \\
Cooling: & telescope 5K, \\
 & detectors 0.1K . \\
\noalign{Wavelengths:}
submillimeter bands: & 0.3, 0.4, 0.5, 0.6, \\
& 0.8, 1, 1.5 mm ; \\
infrared bands: & 3, 10, 30, 100, 200 $\mu$m. \\
\noalign{Detectors:}
in submm bands & bolometer arrays, \\
one-pixel NEP & $10^{-18} W/Hz^{1/2}$, \\
in IR bands & photoconductor arrays , \\
one-pixel NEP & $10^{-17}-3\times 10^{-16} W/Hz^{1/2}$ . \\
\noalign{Sensitivity of the telescope (for integration time 1 s)}
Submm bands: & 3-12 mJy, \\
Infrared bands: & 6-40 mJy. \\
\hline \\
\end{tabular}
\end{table}

\section{Bolometers}
\begin{figure}[ht]
  \begin{center}
    \includegraphics[width=7cm]{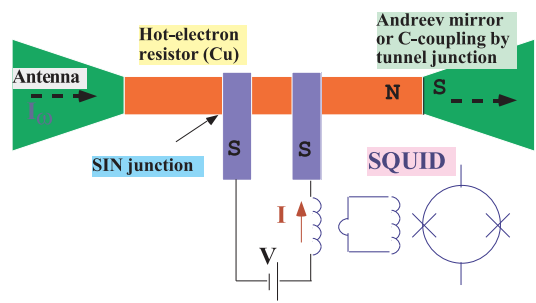}
  \end{center}
\caption{ Schematics of the NHEB bolometer \protect\newline
with SQUID readout.\protect\newline
NHEB - Normal-metal Hot Electron Bolometer,\protect\newline
SQUID - Superconductive Quantum Interferometer Device,\protect\newline
SIN - Superconductor - Insulator - Normal metal }
\label{gromovv_fig:fig4}
\end{figure}

\begin{figure}[ht]
  \begin{center}
    \includegraphics[width=5cm]{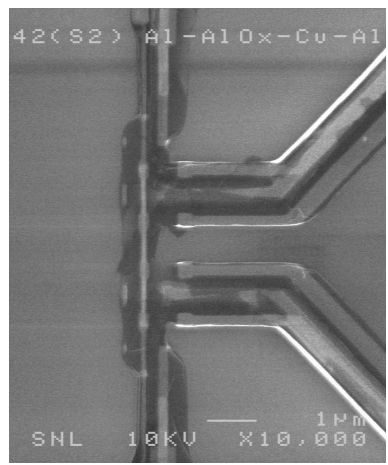}
  \end{center}
\caption{ Sensor element of the NHEB bolometer (electron microscope picture) }
\label{gromovv_fig:fig5}
\end{figure}

High sensitive submillimeter detectors were specially developed for Submillimetron project. The NHEB use
sub\-micron-size sensors cooled to extremely low temperature about 0.1 K to reduce thermal fluctuations to
level less than background emission fluctuations. The initial version of this detector has been proposed
by \cite*{gromovv:NM93}, \cite*{gromovv:N+93}. Figure~\ref{gromovv_fig:fig4}
shows schematics of the bolometer version (\cite{gromovv:GK00})
adapted for array application and low-noise readout with SQUID.
Electromagnetic radiation induces current in planar antenna.
This current heats electron gas in a metal strip.
SIN junction is used as temperature sensor.
The example of the bolometer made in the Microelectronics Center
of Chalmers University is shown in Figure~\ref{gromovv_fig:fig5}.
Measured parameters of this detector are given in Table~\ref{gromovv_tab:tab2}.

\begin{table}[ht]
\caption{ Measured parameters of the NHEB bolometer at temperature 0.1 K }
\label{gromovv_tab:tab2}
\begin{tabular}{ll}
\hline \\
Absorber dimensions & $5\times0.25\times0.07 \mu m^{3}$ \\
Thermal conductance, G= dP/dT & $7\times10^{-14} W/K$ \\
Time constant & $\tau = 5 \mu$s \\
$NEP_{a}$ (experiment, amplifier limited) & $5\times10^{-18} W/Hz^{1/2}$ \\
$NEP_{T}$ (thermal fluctuation noise) & $2\times10^{-19} W/Hz^{1/2}$ \\
\hline \\
\end{tabular}
\end{table}

\section{Discussion}
Other type of bolometer with sensitivity comparable with NHEB is TES (Transition Edge Superconductive)
bolometer, which technology uses experience of TES detectors for X-ray astronomy. Its drawback is
a narrow temperature region of superconductive transition, which exclude functioning of the device
overheated by enhanced background emission possible in sufficient part of submillimeter-wave survey.
On the contrary the NHEB has wider diapason of working temperatures, and uses effect of electronic cooling
improving its characteristics (\cite{gromovv:GK00}).

The orbit of ISS is entirely different from that of most astronomical satellites.
Low, non solar-synchronous orbit put severe restrictions on telescope pointing,
which are nevertheless not so important for sky survey from free flying spacecraft,
where are no problem with shadowing by space station elements.
Protection of cooled optics from strong infrared radiation of Earth and from contamination
(cryocondensation) guaranteed by large reflective cooled screens shown in Figure~\ref{gromovv_fig:fig3}.
The merits of the ISS-related experiment are: it's low cost, reliability and flexibility due to
use of ISS means for deployment, maintenance and instruments change.

New IR/submm projects following on IRAS, COBE and ISO use large mirrors (more than meter), which cannot be
cooled to temperatures lower than tens of K. A development of the Submillimetron project has shown that
progress in observational technique is possible not only by refuse of proven technology of space optics
and cryogenics, but also by progress in detector technology. Full sky data of high sensitive submillimeter
observations with resolution about 1\arcmin~are important complementary to FIRST and Plank missions.

\begin{acknowledgements}

The authors woud like to thank Igor Novikov for useful comments on cosmological objectives
and Tord Claeson for their invaluable support in bolometrs development.
This work was partly supported by grants INTAS 97.731 and ISTC 1239.

\end{acknowledgements}

\end{document}